\documentclass{JINST}
\let\ifpdf\relax
\pdfoutput=1 
\usepackage{graphicx}
\usepackage{subfigure} 
\usepackage{array} 

\makeatletter

\renewcommand{\@thesubfigure}{(\alph{subfigure})\hskip\subfiglabelskip}
\renewcommand{\@@thesubfigure}{(\alph{subfigure})}
\makeatother

\title{The LHCb VERTEX LOCATOR performance and VERTEX LOCATOR upgrade}

\author{Pablo Rodr\'iguez P\'erez$^a$, on behalf of the LHCb VELO Group and the VELO Upgrade group.\\
\llap{$^a$}University of Santiago de Compostela,\\
  Spain\\
  E-mail: \email{pablo.rodriguez@usc.es}}

\abstract{
LHCb is an experiment dedicated to the study of new physics in the decays of beauty and charm hadrons at the Large Hadron Collider (LHC) at CERN.  
The Vertex Locator (VELO) is the silicon detector surrounding the LHCb interaction point.
The detector operates in a severe and highly non-uniform radiation environment.
The small pitch and analogue readout result in a best single hit precision of 4 $\rm \mu$m.

The upgrade of the LHCb experiment, planned for 2018, will transform the entire readout to a trigger-less system operating at 40 MHz event rate. 
The vertex detector will have to cope with radiation levels up to 10$^{16}$ 1 MeV$\rm n_{eq}/cm^2$, more than an order of magnitude higher than those expected at the current experiment. 
A solution is under development with a pixel detector, based on the Timepix/Medipix family of chips with 55 x 55 $\rm \mu m$ pixels. 
In addition a micro-strip solution is also under development, with finer pitch, higher granularity and lower mass than the current detector. 
The current status of the VELO will be described together with recent testbeam results.
}


\begin{document}

\section{The VELO detector at the LHCb experiment}

  \subsection{The LHCb experiment}
  LHCb \cite{lhcb} is an experiment with forward geometry and has shown excellent vertex, momentum and particle identification capabilities.
  LHCb has recorded for analysis 1.0 $\rm fb^{-1}$ pp collisions at 7 TeV center of mass energy during 2011. 
  After 2011-2012 winter shutdown the energy was increased to 8 TeV, and since then, 1.4 $\rm fb^{-1}$ have been recorded (September 2012).
  Currently LHCb is running at a luminosity of $\rm \mathcal{L}= 4\times10^{32}cm^{-2}s^{-1}$ with an average pile up of 1.4 interactions per bunch crossing.
  Both values are above design values, which is an indication of the excellent performance of the experiment.
 
  \subsection{The VERTEX LOCATOR (VELO)}
  
  \begin{figure}[tb]
  \centering \includegraphics[width=0.47\textwidth]{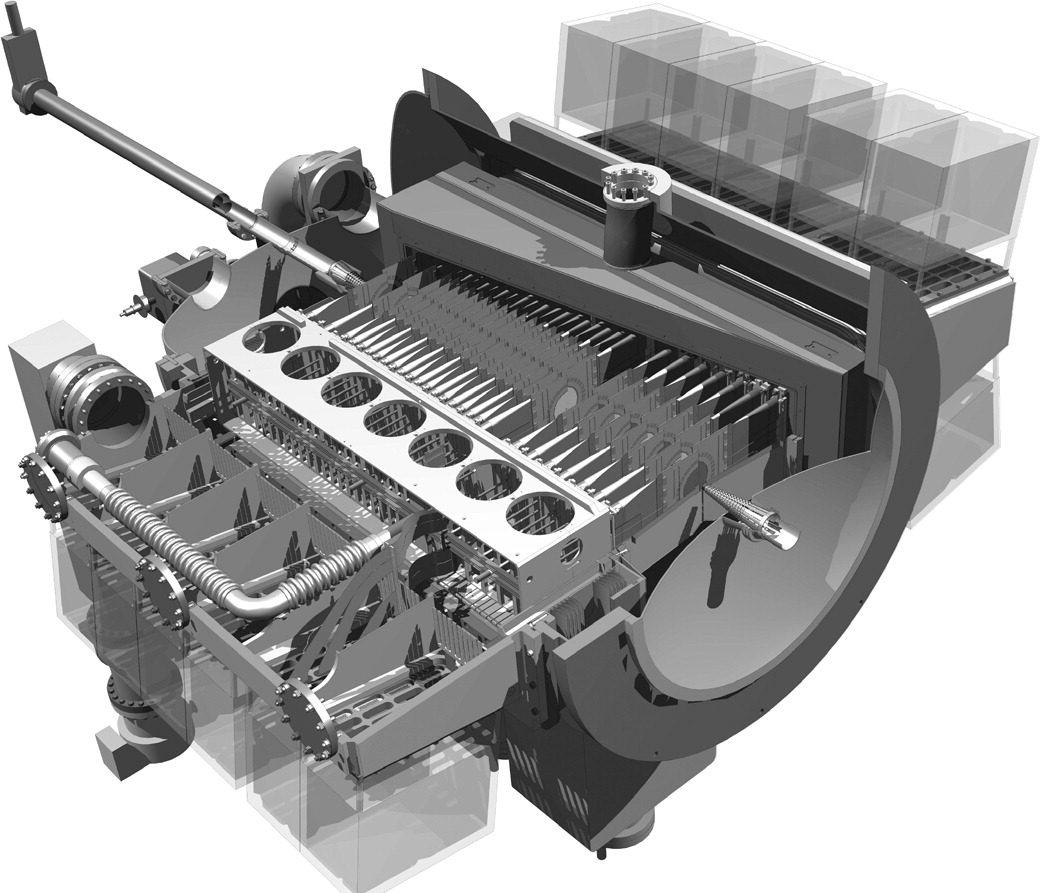}
  \caption{\small CAD image of the VELO detector. The beam pass through the center of the detector where the collisions occur.}
  \label{fig:velo_modules}
  \end{figure}
  
  The Vertex Locator (VELO) \cite{vanBeuzekom200821} is the silicon detector surrounding the LHCb interaction point.
  It is located only 7 mm from the beams during normal operation and measures very precisely the primary vertex and the decay vertices.
  It consists of two retractable detector halves with 21 silicon micro-strip tracking modules each (figure~\ref{fig:velo_modules}). 
  A module is composed of two $\rm n^+$-on-n 300~$\rm \mu$m thick half disc sensors with R and $\rm \phi$-measuring micro-strip geometry. 
  The detector is operated in vacuum and a bi-phase CO$_{2}$ cooling system is used. 
  The sensors are readout with the analogue front-end chip (Beetle \cite{beetle}).
  
  Only when the stable beam flag is raised both halves of the VELO are closed, allowing a small overlap in the inner region, reaching a distance of only 7 mm to the beam line. 
  
  \subsection{VELO performance}
\begin{figure}[tb]
  \centering
  \subfigure[Point Resolution]{
	  \includegraphics[width=0.4\textwidth]{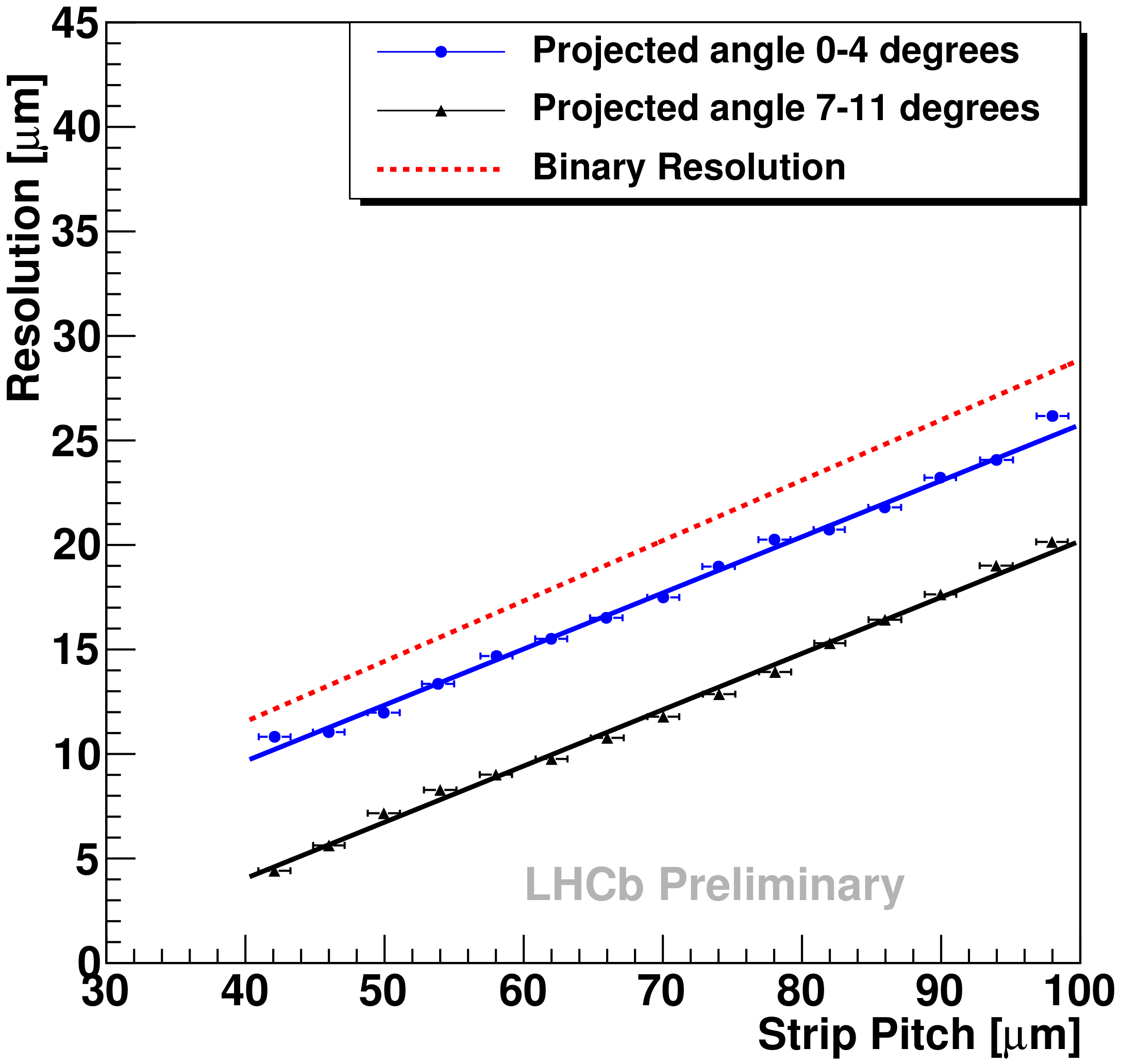}
	  \label{fig:subfig1}
  }  
  \subfigure[Primary Vertex]{
	  \includegraphics[width=0.55\textwidth]{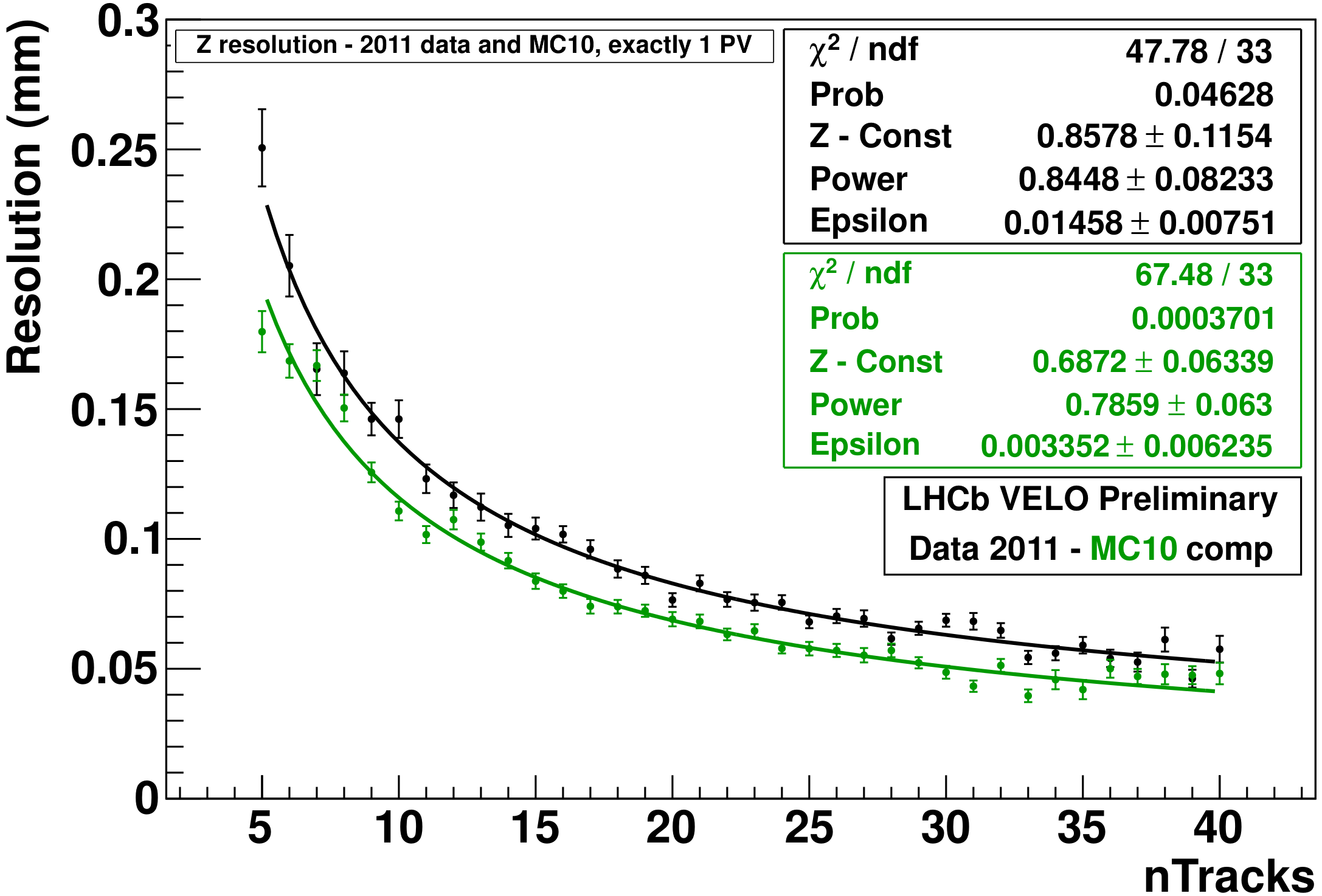}
	  \label{fig:subfig2}
  }
  \subfigure[Impact Parameter]{
	  \includegraphics[width=0.4\textwidth]{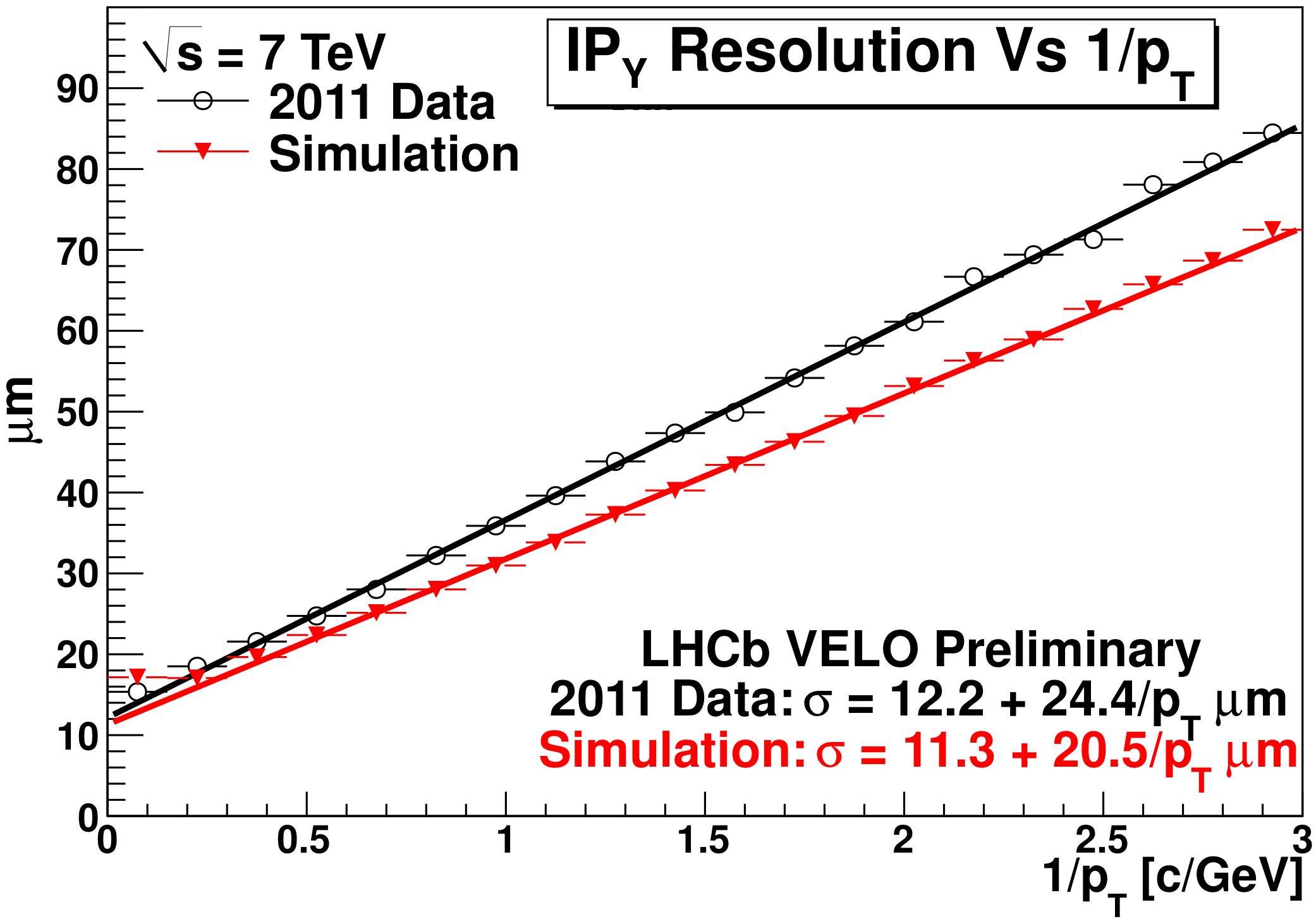}
	  \label{fig:subfig3}
  }
  \caption[Optional caption for list of figures]{Figure \subref{fig:subfig1} shows the single hit resolution of R-sensors as function of strip pitch for two different projected angles. The binary resolution is shown as reference. Figure \subref{fig:subfig2} shows the vertex resolution along the beam axes (Z) as function of the number of tracks. Figure \subref{fig:subfig3} shows the impact parameter at optimal angle.}
  \label{fig:subfigureExample}
  \end{figure}

  The 1 MHz L0 trigger rate is reduced to a few kHz by full event reconstruction in the computer farm.
  The VELO plays a crucial role providing vertexing information for this decision.
  On the tracking and vertexing side, the VELO provides accurate reconstruction of primary and decay vertices, reaching a point resolution of 4~$\rm \mu$m in the 40~$\rm \mu$m pitch region at optimal angle ($\rm \sim 10^o$, figure~\ref{fig:subfig1}). 
  The primary vertex resolution (PV) is 13~$\rm \mu$m in the longitudinal direction for about 25 tracks (figure~\ref{fig:subfig2}). 
  The final momentum dependent impact parameter resolution (IP) is $\rm 11.5+24.5/p_T$, where $\rm p_T$ is the transverse momentum in GeV/c (figure~\ref{fig:subfig3}).
  Accurate determination of the IP is very important for LHCb.
  The relationship between the IP and the PV for a particle is used in the trigger as cut in many physics analysis.
  
  As result, a lifetime resolution of $\rm \sim$ 50 fs was achieved for the flagship channel $\rm B_s \rightarrow J/\psi\phi$ \cite{Krocker:1404788}, typical of those channels where the resolution is necessary to resolve fast $\rm B_s$ oscillations. Excellent resolution is also very important for background suppression in rare decay channels.
     
  \subsection{Radiation damage}
\begin{figure}[tb]
  \centering 
  \subfigure[Leakage current versus delivered luminosity]{
    \includegraphics[width=0.5\textwidth]{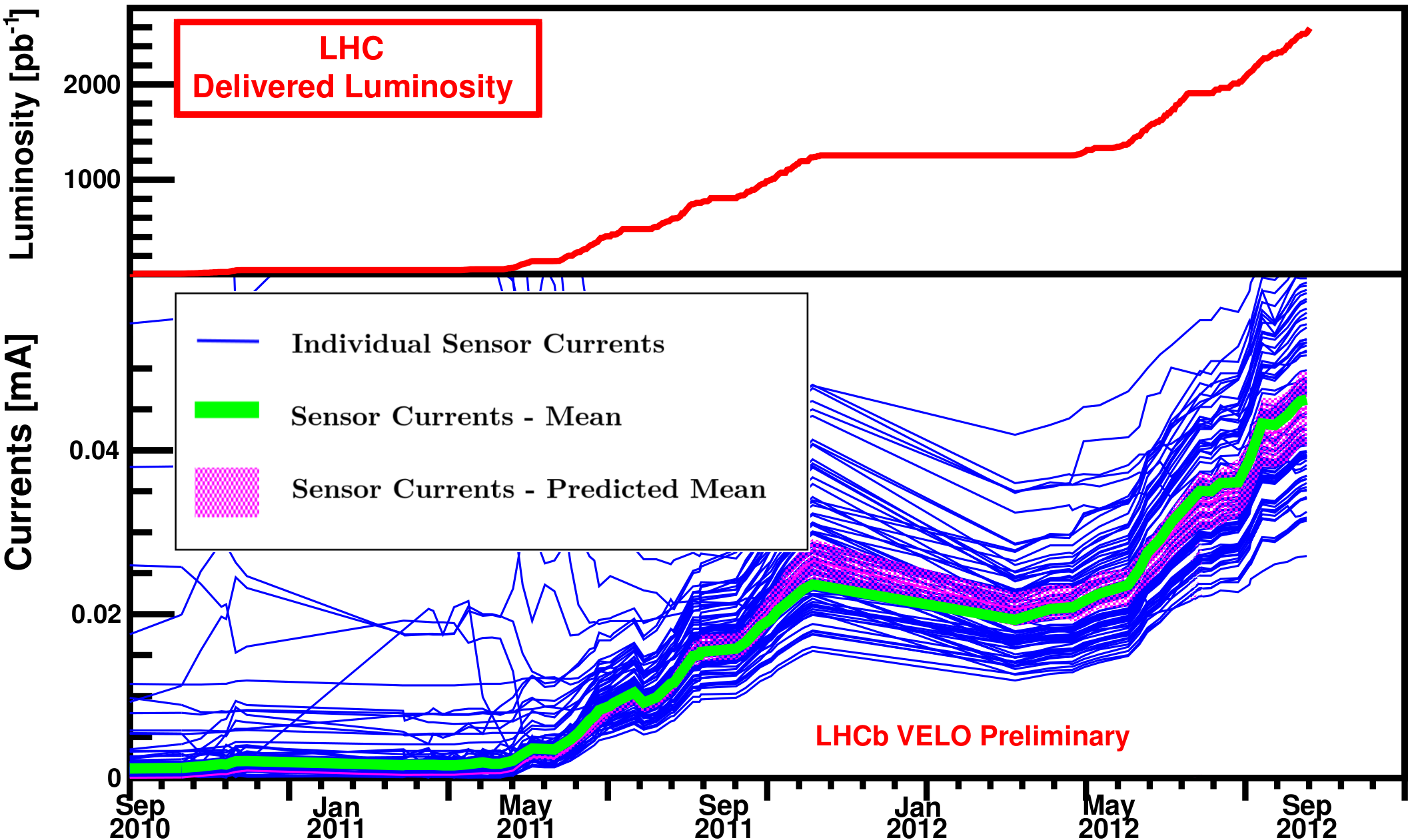} 
    \label{fig:leakage_current}
  }
  \subfigure[Effective Depletion Voltage (EDV)]{
	  \includegraphics[width=0.45\textwidth]{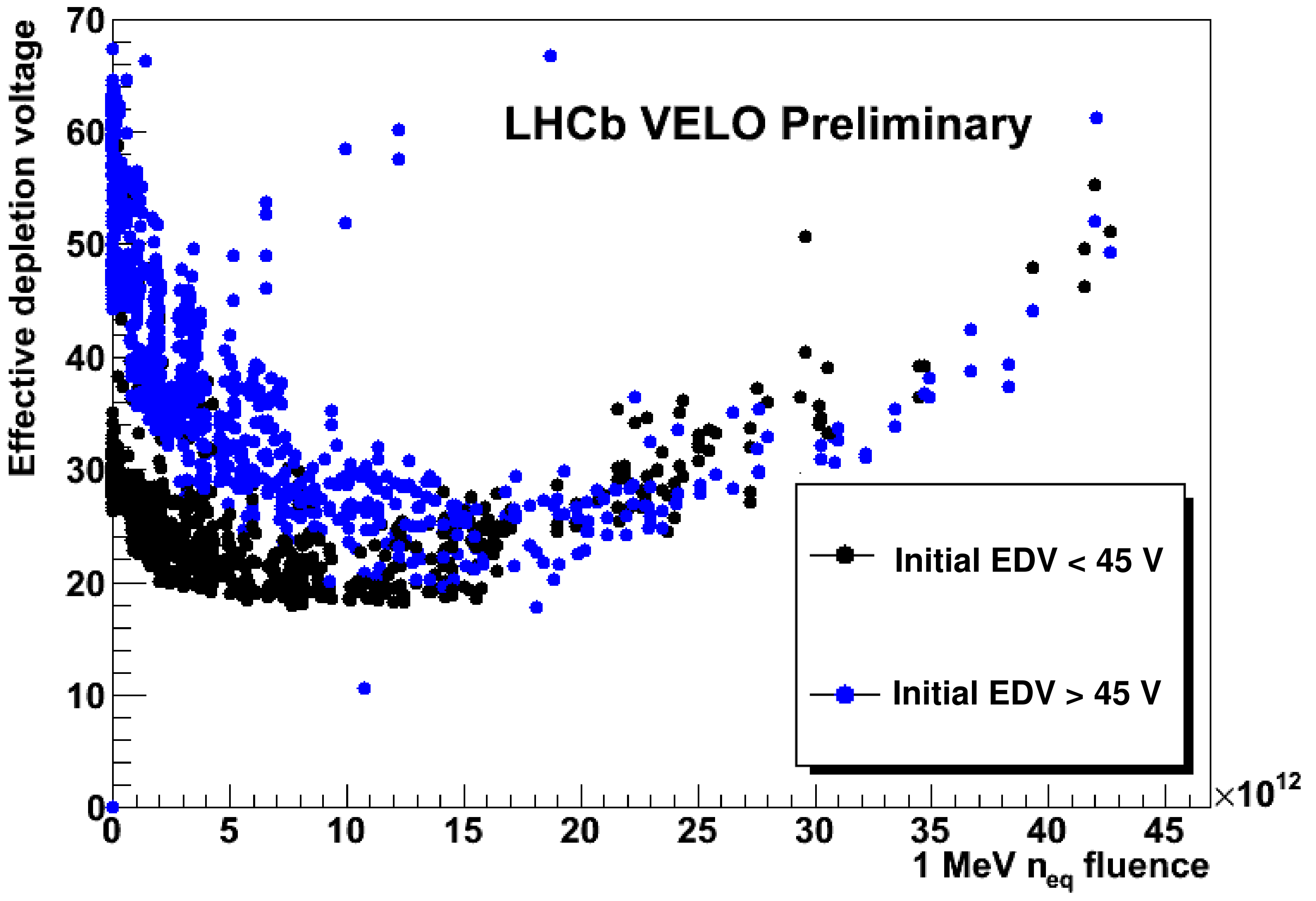}
	  \label{fig:edv}
  }
  \caption[]{Radiation effects on VELO. Left plot shows the measured leakage current at $\rm -8 ^oC$ as function of the luminosity. Each blue line corresponds to a single sensor, the green curve is the mean current value excluding those sensors with initial high currents, and the pink band corresponds to predicted currents. The effective depletion voltage (EDV) for different sensors is shown in left plot. Blue dots corresponds to sensors with initially lower leakage currents, and black to sensors with high leakage current at the beginning.}
  \label{fig:rad_damage}
  \end{figure}

  Due to the small distance to the interaction point severe radiation damage is foreseen in the VELO sensors \cite{velo_rad_damage}, with doses $\rm \sim0.6\times10^{14} MeV n_{eq} cm^{-2}$ per accumulated $\rm fb^{-1}$. 
  The increase of the leakage current was measured as a function of the delivered luminosity (see figure~\ref{fig:leakage_current}).
  The periods with a decrease in the current corresponds to the annealing during shutdown periods or to occasional detector warm ups.
  The typical current increase is around 1.9~$\rm \mu$A per 100~$\rm pb^{-1}$ delivered.
  The effective depletion voltage is measured with HV scans during dedicated datataking periods. 
  Figure ~\ref{fig:edv} displays the results of these scans split by detector region and received dose.
  Most of the innermost parts of the $\rm n^+$-on-n sensors have already passed through type inversion.
  Studies of the cluster finding efficiency have revealed a radiation induced charge loss in R-type sensors. 
  For these sensors the mean of the collected charge is lower, and the distribution has experienced deformation due to a growing fraction of clusters with very small charge.
  The effect has been linked to coupling between routing lines in the second metal layer and the aluminium of the strips in the first metal layer. 
  After irradiation, phantom signals in the inner strips (with readout through lines in the second metal layer) have been detected when the particles cross the outer strips.
  In the case of the $\rm \phi$ sensors the effect is mitigated because lines in both metalizations are collinear.
  The exact mechanism and the relationship with radiation damage is under study.
  For more information about radiation damage in the VELO, see \cite{Hickling:1392353, rad_damage_poster}.

\section{The VELO upgrade}
  \subsection{The LHCb upgrade}
  The LHCb experiment will record around 1-2~$\rm fb^{-1}$ of data per year until end 2017, accumulating in excess of 5~$\rm fb^{-1}$.
  Afterwards many of the physics channels will still be statistically limited. 
  The LHC is already capable of delivering much higher luminosity to the experiment than the current one, therefore the LHCb has planned an upgrade for 2018 which is independent, but compatible with, the LHC upgrade.
  After the upgrade the experiment will operate at luminosities up to 2~$\rm \times10^{33} cm^{-2}s^{-1}$ and will readout collisions at 40 MHz. 
  In these new conditions we expect to collect more than 5~$\rm fb^{-1}$ per year, with improvements of up to a factor 2 in the efficiencies in hadronic channels.
  
   \subsection{The VELO upgrade} \label{sec:velo_upgrade}
  It is a big challenge for the VELO to cope with the requirements of the LHCb upgrade, specially in terms of radiation damage and data bandwidth  \cite{Collins2011S185, nim_velo_upgrade}.
  These new conditions must be satisfied keeping, or improving, the performance of the current VELO in terms of precision, pattern recognition and impact parameter resolution. 
  Many systems of the current VELO detector, such as the motion system, the vacuum, the power systems (LV/HV) and the evaporative $\rm CO_2$ cooling system, will be reused.

\begin{figure}[b]
  \subfigure[]{
    \centering \includegraphics[width=0.47\textwidth]{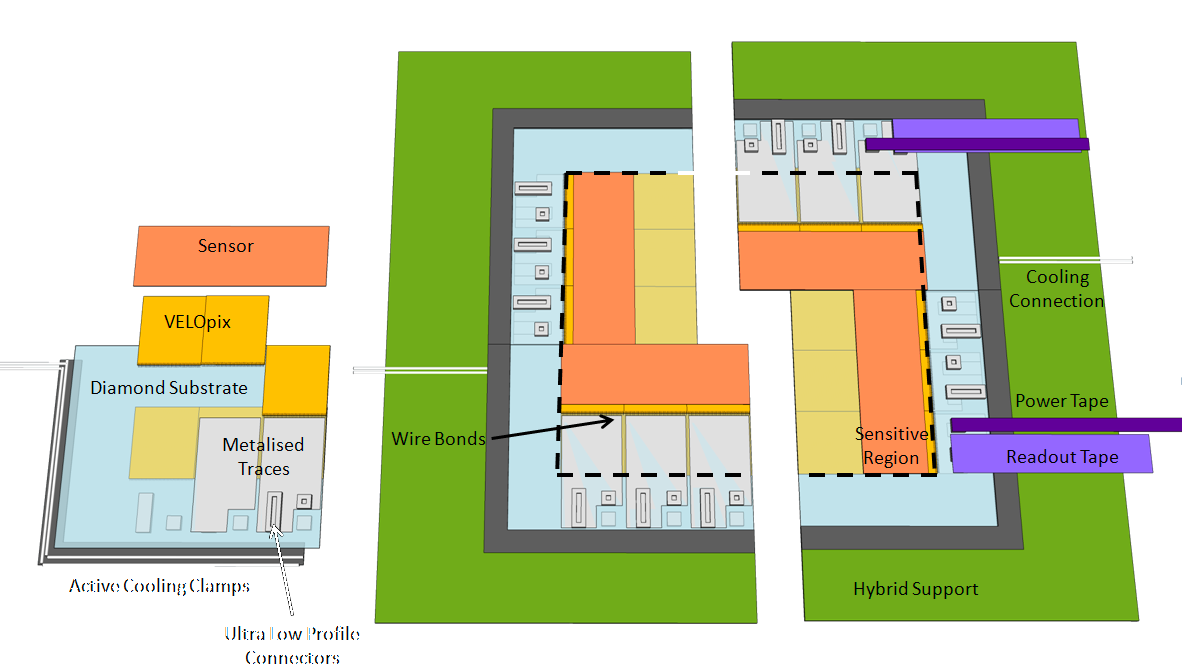} 
    \label{fig:pixel_hybrid}
  }
  \subfigure[]{
    \centering \includegraphics[width=0.42\textwidth]{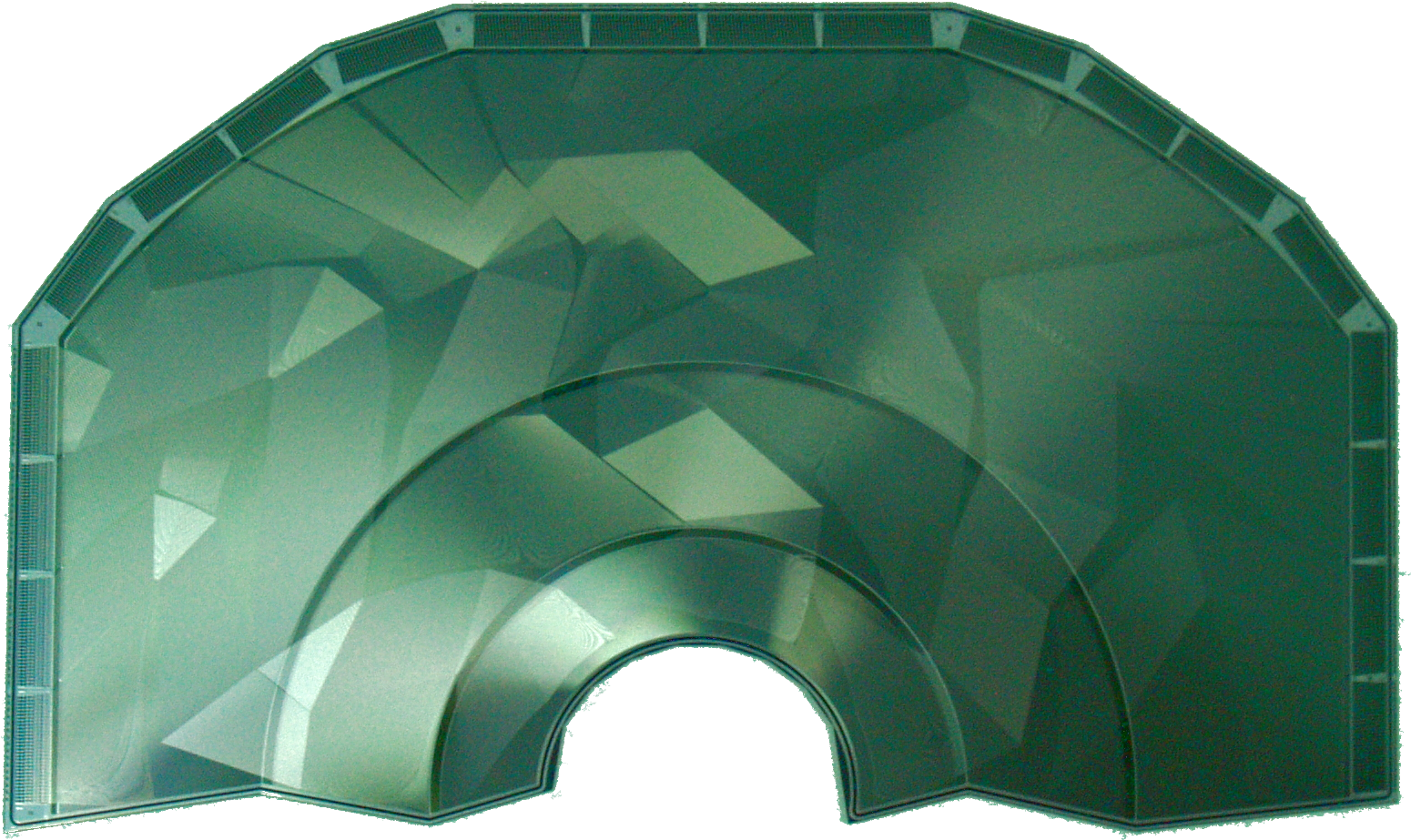}
    \label{fig:strip}
  }
  \caption[]{\small L-shape module for the pixel prototype \ref{fig:pixel_hybrid}, and prototype of $\rm \phi$ strip sensor produced by Hamamatsu  \ref{fig:strip}}
  \label{fig:velo_prototypes}
  \end{figure}

  On the sensor side two options are under consideration, pixel sensors with an ASIC based on Medipix/Timepix family (figure~\ref{fig:pixel_hybrid}), and strips sensors with a new and more advanced ASIC (figure~\ref{fig:strip}).
  Both options imply a completely new module and RF Foil design.
 
    \paragraph{Requirements}
    One of the main constraints for the VELO upgrade is the bandwidth associated to a triggerless experiment, specially with the increase of the luminosity.
    The total data rate coming out from the detector will be around 2.4 Tbit/s for the strip option, or 2.8 Tbit/s for the pixel option.
    The other main issue is the radiation damage, as the upgraded VELO must be able to withstand radiation levels above 400~MRad or $\rm 10^{	16}$ 1 MeV $\rm n_{eq}/cm^2$.

    \paragraph{FE Electronics}
    \subparagraph{Pixel option}
    The baseline choice for the upgraded VELO is a pixel ASIC called Velopix, which is based on the Timepix3 \cite{Llopart2007485}.
    Some of the Velopix characteristics are summarized in Table \ref{tab:velopix_features}.
    The ASIC will integrate local intelligence at the pixel level for time-over-threshold measurement and sparse readout, on-chip zero suppression and packet-based readout.
    In addition, Velopix must be immune to single-event upsets in its digital logic. 
    The main difference between TimePix3 and Velopix is that the latter should cope with higher particle flux, hence the bandwidth must be 6 or 7 times greater.
    The Medipix3 family was designed to be radiation hard, and the Velopix chip will inherit this feature to cope with radiation conditions of the upgraded VELO.

    \subparagraph{Strips option} 
      The backup solution is the strip sensor with an enhanced design and readout chip.
      The development of the ASIC is ongoing in synergy with other silicon detectors in LHCb.
      It will implement embedded solutions such as clustering, sparsification, common mode suppression and pedestal subtraction.
      A strip design results in a lower material budget than a pixel based solution.
                
    \begin{table}
      \centering
            \caption{Main Velopix features} 
	  \begin{tabular}{|>{\bfseries}l| l|}
	  \hline
	  Pixel array & 256 x 256  \\
	  \hline
	  Pixel size & 55 $\rm \mu$m $\rm \times$ 55$\rm \mu m$ \\  
	  \hline  
	  Minimum threshold &  $\rm \sim$ 500 $\rm e^-$ \\
	  \hline
	  Peaking time &  < 25 ns \\  
	  \hline  
	  Time walk &  < 25 ns \\
	  \hline 
	  Measurements & Time-of-Arrival \& Time-over-Threshold \\  
	  \hline
	  ToT range &  4 bit (1-2 MIP) \\
	  \hline
	  ToA resolution / range &  25 ns / 12 bit \\  
	  \hline  
	  Count rate &  Up to 500 Mhit/s/chip \\
	  \hline
	  Readout &  Continuous, sparse, on-chip clustering \\  
	  \hline  
	  Output bandwidth & > 12 Gbit/s \\
	  \hline
	  Power consumption & < 3 W/chip \\  
	  \hline  
	  Radiation hardness & > 500 MRad, SEU tolerant \\ 
	  \hline
	  \end{tabular}        
      \label{tab:velopix_features}
    \end{table}
    
    \paragraph{Sensors}
    The baseline option for the sensor is planar silicon with $\rm n^+$ type readout \cite{Casse2010401}. 
    The challenges for both pixels and strips include coping with a large and very non uniform radiation dose, reducing the material budget and incorporating as narrow as possible guard rings.
    
    For the strip option, sensors should have a minimum pitch of 30~$\rm \mu$m or below (currently 40~$\rm \mu$m), 200~$\rm \mu$m thickness (currently 300~$\rm \mu$m), \emph{n-on-p} type, and with a variable pitch in order to keep the same occupancy per strip. 
    For this option, several prototypes with R and $\rm \phi$ design where produced by Hamamatsu, and are currently under characterization in the University of Santiago de Compostela.
    
    In the case of the pixel option, other alternatives than planar sensors are under consideration, like 3D sensors \cite{Pellegrini2012} and diamond sensors. Several 3D sensors from CNM were studied during last testbeam campaigns.

    \paragraph{Mechanics and cooling}
    As it was explained in section \ref{sec:velo_upgrade} the cooling system for the VELO upgrade will be based on the same evaporative $\rm CO_2$ system that is installed in the current VELO.
    However, two main techniques are being investigated in order to cool the module sensors. 
    
\begin{figure}[tb]
    \centering
    \subfigure[CVD Diamond support and TPG block (top) and Micro-channel (bottom)]{
	    \includegraphics[width=0.47\textwidth]{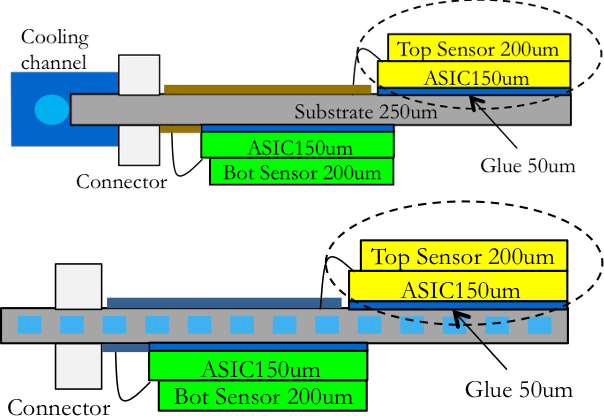}
	    \label{fig:cooling1}
    }  
    \subfigure[Micro-channel Cooling: substrate outlet]{
	    \includegraphics[width=0.4\textwidth]{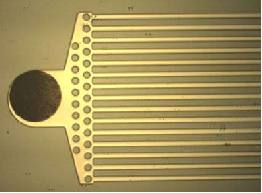}
	    \label{fig:microchannel}
    }
    \caption[]{The two options under consideration for module cooling. Figure \subref{fig:cooling1} (top) is the baseline solution based in a diamond support. Figures \subref{fig:cooling1} (bottom) and \subref{fig:microchannel} show the alternative solution based on microchannels etched into the silicon module's support. Microchannels dimensions are  200~$\rm \mu m \times 70 \mu m$.}
    \label{fig:cooling}
    \end{figure}

      \subparagraph{CVD Diamond}
      A proposed solution (figure~\ref{fig:cooling1} top), and consists of a chemical vapor deposition (CVD) diamond support plus a thermal pyrolitic (TPG) cooling block. 
      The diamond support guarantees a high thermal conductivity, highly robust and low material budget.
      For this solution the R\&D is currently centered on diamond metalization.
      
      \subparagraph{Micro-channel}
      Through etching processes, a set of micro-channels are developed into a silicon substrate \cite{microchannels, 1748-0221-7-01-C01111}.
      The coolant ($\rm CO_2$) is forced  to circulate through them at a pressure of $\rm \sim$~60~bar (figures \ref{fig:cooling1}-bottom and \ref{fig:microchannel}). 
      It is an attractive solution because the channel layout can be adapted to the heat dissipation needs.

    \paragraph{RF-Foil}
    The RF-Foil is the box that encloses the detector halves in a secondary vacuum.
    It is customized for the sensor layout to allow the placement of sensitive areas at the required distance from the beams, while slightly	 overlapping in closed configuration.
    The production of the new RF-Foil is ongoing with milling techniques, and a prototype has been produced for L-shape modules, which is the module for the pixel option. 
    A large R\&D is under way to manufacture a full scale prototype within the required specifications.

  \subsection{The testbeam program}
  
  We have built a testbeam telescope with 9 Timepix sensors \cite{timepix_paper}, 8 in ToT\footnote{\textbf{T}ime \textbf{o}ver \textbf{T}hreshold provides a value proportional to the deposited energy.} configuration and the last one in ToA\footnote{\textbf{T}ime \textbf{o}f \textbf{A}rrival provides the time-stamping of the track}.
  This telescope has a point resolution at the Device Under Test (DUT) level of $\rm \sim$  1.5~$\rm \mu$m, a time tag resolution of $\rm \sim$ 1 ns and handles a track rate $\rm \approx$ 10 kHz.
  The Timepix hybrids are connected to an innovative system called RELAXed which is based on FPGAs and can read 4 hybrids in parallel sending the data through an Ethernet connection.
  
\begin{figure}[tb]
    \centering
    \subfigure[Residuals of different planar sensors]{
	    \includegraphics[width=0.45\textwidth]{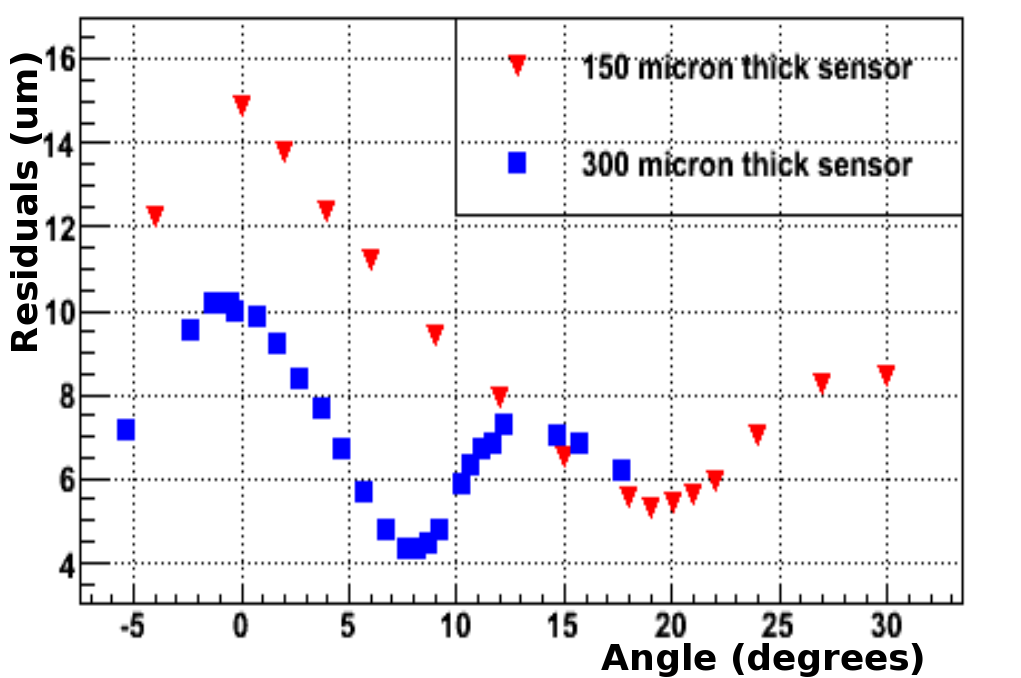}
	    \label{fig:testbeam_a}
    }  
    \subfigure[Efficiencies in column region measured for a 3D sensor]{
	    \includegraphics[width=0.45\textwidth]{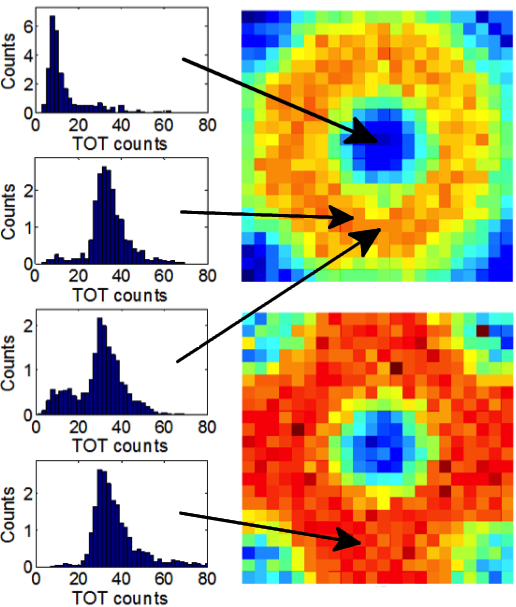}
	    \label{fig:testbeam_b}
    }
    \caption[]{Figure \subref{fig:testbeam_a} shows a comparison between two planar sensors of different thickness. Figure \subref{fig:testbeam_b} shows the energy collected for a 3D sensor in a column neighborhood.}
    \label{fig:testbeam}
  \end{figure}

  Up to 8 different DUTs were analyzed with this telescope such as irradiated Medipix2 chips, thinned planar sensors (150~$\rm \mu$m thick) or 3D sensors \cite{3D_plots}. 
  In figure ~\ref{fig:testbeam_a} a comparison between two planar sensors of 150~$\rm \mu$m and 300~$\rm \mu$m thickness is shown. 
  The residual value inform us about the resolution of the reconstructed cluster in the DUT with respect to the extrapolated track given by the telescope, and it changes with the angle and the thickness as expected.
  In 3D sensors polarization is made between n-type and p-type columns etched into the silicon.
  Due to this particular geometry, a certain loss of efficiency is expected when a particle crosses the sensor trough a polarization column, as it can be seen in figure~\ref{fig:testbeam_b}.

\section{Conclusions}
The VELO is operating smoothly under conditions beyond its design.
It is a key element in the physics output of the LHCb due to its exceptional performance in terms of vertex and impact parameter resolutions.

As a consequence of the high radiation doses, the VELO sensors have already exhibited radiation damage, and type inversion is observed in the inner regions of the sensors.
Nevertheless, the impact in the performance of the detector is not significant up to date.

In 2018 an upgrade of the VELO is planned that should be able to cope with the new requirements of the upgraded LHCb in terms of radiation hardness and readout bandwidth.

Some R\&D is still needed in order to select the sensor technology and cooling for the upgraded detector.
The candidates are pixels or strips with their associated FE electronics for the sensor and metalized diamond or microchannel for the cooling. 
The goal is to have confident technologies before 2014 when the production stage will start.

\bibliographystyle{JHEP}
\bibliography{Velo}

\end{document}